\documentclass[12pt,preprint]{aastex}
\newcommand {\etal} {{\it et~al.}}

\newcommand {\ergcms}{ergs cm$^{-2}$ s$^{-1}$}
\newcommand {\ergs}{ergs s$^{-1}$}
\newcommand {\go} {\mathrel{\hbox{\rlap{\lower.55ex \hbox {$\sim$}}
           \kern-.3em \raise.4ex \hbox{$>$}}}}
\newcommand {\lo} {\mathrel{\hbox{\rlap{\lower.55ex \hbox {$\sim$}}
           \kern-.3em \raise.4ex \hbox{$<$}}}}

\begin{document}

\title{AN XMM-NEWTON OBSERVATION OF 4U1755-33 IN QUIESCENCE: 
EVIDENCE FOR A FOSSIL X-RAY JET}
\author{Lorella Angelini\altaffilmark{1} and Nicholas E. White}

\affil{Laboratory for High Energy Astrophysics, Code 660, 
NASA Goddard Space Flight Center, Greenbelt, MD 20771}
\email{angelini@davide.gsfc.nasa.gov; nwhite@lheapop.gsfc.nasa.gov}

\altaffiltext{1}{Universities Space Research Association}

%\lefthead{ Angelini \& White}
%\righthead{4U1755-33 in Quiescence}

\begin{abstract}
We report an XMM-Newton observation of the Low mass X-ray Binary (LMXB) and
 black hole candidate 4U1755-33. This source had been a bright persistent source for at least 25 yrs, but in 1995 
entered an extended quiescent phase.  4U1755-33 was not 
detected with an upper limit to the 2--10 keV luminosity 
of  $5 \times 10^{31}d_{4kpc}^2$ \ergs (where $d_{4kpc}$ is the distance in units of 4 kpc) -- 
consistent with the luminosity of other black hole candidates in a quiescent state. 
An unexpected result is the discovery of a narrow $7' $ long X-ray jet-like feature
 centered on the position of 4U1755-33. 
The spectrum of the jet is similar to that observed from other galactic and extragalactic 
jets thought to be associated with  accreting black holes.  The detection of a fossil jet provides 
additional evidence supporting the black hole candidacy of 4U1755-33. 
The spectral properties of three bright serendipitous sources in the field are reported and 
it is suggested these are background active galactic nuclei (AGN).

\end{abstract}


\section{Introduction}

The LMXB 4U1755-33 had been a permanent feature of the X-ray sky 
since its discovery in the first all-sky X-ray survey by the Uhuru 
satellite in 1970 (Giacconi {\it et al} 1974), with a typical flux of $\sim 100$ $\mu$Jy. 
An observation by 
{\it RXTE} in January 1996 revealed that 4U1755-33 had turned off, 
with an upper limit to a flux of $\leq$ 1 $\mu$Jy (Roberts \etal\ 1996). 
The {\it RXTE} all sky monitor shows it  has been quiescent since then, to the time of writing.  
4U1755-33 has an orbital period of 4.4 hr, which
was established from regular periodic dipping in the X-ray lightcurve 
(White \etal\ 1984), and confirmed in photometric variations in the 
18--19 magnitude optical counterpart V4143 Sagittarii (Mason, Parmar \& White 1985). 
Since the X-ray source turned off, the optical counterpart  became 
fainter than V $>$ 22 (Wachter \& Smale 1998). 

When it was active 4U1755-33 was noted for having an ultrasoft X-ray  
spectrum, similar to that of black hole candidates (White \& 
Marshall 1984, White \etal\ 1984). The spectral evidence for the 
compact object being a black hole was further strengthened when broad band 
observations revealed a hard X-ray tail, again similar to that seen in other
 black hole systems (Pan \etal\ 1995). However, the 
possibility that a neutron star might mimic this spectral signature 
has not been ruled out (see e.g. Seon \etal\ 1995 and references 
therein).

The study of LMXBs during quiescence, when  
the luminosity is a factor of 1 million or more fainter than the peak, 
has become an important tool to investigate the nature of the compact object 
and the physics of the accretion process (see e.g. Garcia \etal\ 2000). 
Observations with ASCA, Beppo-SAX and ROSAT were the first to detect faint 
emission from these LMXBs in quiescence with a luminosity of typically 
$\le 10^{32}$ \ergs\  with fluxes of $\le 10^{-14}$ \ergcms\ 
(e.g. Menou \etal\ 1999). 
 The luminosity of systems thought to contain a black hole appears to be significantly 
fainter than systems thought to contain a neutron star (e.g. Menou \etal\ 1999). 
This lower luminosity of the black hole systems has been attributed as due to the lack 
of a solid surface on the compact object and used as supporting the existence of the
 black hole event horizon (e.g. Garcia  \etal\ 2000). 
 These systems are also used to test models for the accretion disk emission 
e.g. the Advection Dominated Accretion Flow (ADAF) model (Narayan, McClintock \& Yi 1996).

With the advent of XMM-Newton and 
{\it Chandra} it is possible to extend these studies using more sensitive observatories
(e.g. Hameury {\it et al} 2002).
In this paper we present the results of an observation of 4U1755-33 in its quiescent 
state using the European Space Agency's XMM-Newton Observatory 
(Jansen {\it et al} 2001).

\section{Results}
\subsection{The Observation}

4U1755-33 was observed by the XMM-Newton observatory on March 3, 2001 
starting at 05:13 hr UT and ending at 10:30 hr. XMM-Newton has three X-ray 
telescopes with two different types of CCD arrays in each focal plane. 
Two CCD cameras use MOS arrays and one a PN array. 
The data were processed using XMM-Newton Science Analysis 
System (SAS) version 5.3.3. 
Images were accumulated from the cleaned event files selecting 
good events as recommended in the users guide. 
Images were made for each CCD camera in the 0.5--4.5 keV band, 
6.2--6.8 keV iron band and 0.5--10 keV band. 
The good time exposure in each camera was: 
MOS-1=18127s, MOS-2=18107s and PN=15359s.

The three images in the 0.5--4.5 keV band were summed together and the 
total image is shown in Figure 1. The position of 4U1755-33 is at the 
center of the image, and is marked with a circle. While many sources are seen in the image, 
there is none at the position of 4U1755-33. 
A striking aspect of the image is a $\sim 7 ' $ long jet-like 
feature of extended emission that passes through the position of 4U1755-33. 
There are several  
bright spots embedded in the extended emission, in particular on the southern side. 

To examine the region around 4U1755-33 in more detail in Figure 2 we 
show the individual images from the PN camera and the summed MOS cameras
 for the 0.5--10 keV band. The lower two images zoom in on the position 
of 4U1755-33. We separate the MOS and PN cameras so that the gaps in the 
image caused by CCD chip boundaries or bad columns can be seen. 
The position of 4U1755-33 is marked on the PN image.

\subsection{4U1755-33 in Quiescence}

An upper limit to the flux from 4U1755-33 was obtained by selecting an 
optimum sized region around the source. A circle of radius 7 arc sec was 
used to minimize the contribution from the extended emission, this includes 
about 45\% of the telescope point spread function (PSF). Because of the 
different energy response of the two types of CCD cameras and different coverage 
across the CCD we separately 
estimated an upper limit for the summed MOS and the PN images. By taking 
many different backgrounds across the image an average background count rate 
was estimated and found to be similar to the count rate at the position of 
4U1755-33. 

A three sigma upper limit was derived from the count rate in the 
source box using the statistics method described by Gehrels (1986).  
The upper limits derived for the summed MOS and PN are 
$ < 3.4\times 10^{-4}$  count/s and  $ < 1.3\times 10^{-3}$ count/s in 
the 0.5--2 keV and $ < 3.5\times 10^{-4}$ count/s and 
$ < 1.4\times 10^{-3}$ count/s in the 2--10 keV bands. 
This was then corrected by a factor of 2.2 to correct for the portion 
of the telescope point spread function not included in the circle.

The corresponding flux level depends on the assumed spectrum. 
When active the spectrum of 4U1755-33 had a characteristic temperature 
of $\sim$ 2 keV and an equivalent hydrogen column density of 
$5 \times 10^{21}$ cm$^{-2}$ (White \etal\ 1984). 
The distance to 4U1755-33 is very uncertain with estimates ranging 
from 4 to 9 kpc (see Wachter and Smale 1998). Table 1 summarizes the 
various flux upper limits for the MOS and PN, using a bremsstrahlung
model for  two temperatures of 2 and 10 keV. The latter temperature is 
representative of other transient black hole systems detected in quiescence 
(e.g. Narayan \etal\ 1996). These upper limits are corrected for any attenuation by 
interstellar absorption. The MOS gives the most restrictive upper limits, 
which for the 2--10 keV band corresponds to $3.6 \times 10^{31}d_{4kpc}^2$ \ergs\ 
(where $d_{4kpc}^2$ is the distance in units of  4 kpc) assuming a 10 keV spectrum. 

The ratios of the quiescent luminosity to the peak luminosity and the Eddington 
luminosity are other useful parameters to compare the relative activity across systems 
(e.g. Menou \etal\ 1999). This has been applied to transient systems, 
where there is well defined peak luminosity. 
For a steady source such as 4U1755-33 it is less clear that this is a meaningful parameter.

\subsection{Extended Emission}

The extended streak of emission centered on 4U1755-33 is suggestive of a 
two-sided jet ejected from this X-ray binary when it was active. 
Profiles along the extended emission were made by slicing the MOS image along 
the jet-like feature, and for comparison also along an offset position. 
These are shown in Figure 3 (the regions used are indicated 
in Figure 2). The bright point source seen passing through the lower slice 
gives a reference as to the size of the XMM-Newton telescope point spread 
function (PSF). The bright knots on the left side of 4U1755-33 appear somewhat 
broader than the PSF of the telescope, suggesting that they are not point sources.

Spectra were extracted from the MOS detectors for two different sized regions 
along the extended feature, centered on the position of 4U 1755-33, as 
indicated in Figure 2. A background spectrum was accumulated from a source 
free region parallel to the jet. The summed MOS spectra were fit to a number of 
different models: power law, thermal bremsstralung and a MEKAL plasma model. 
The statistics are poor and an acceptable fit is found for all three models.  
The PN camera has less uniform coverage because of gaps between the CCDs and 
bad columns, and was not used for the larger region.

For the inner region the best fit is a photon index of $1.9 \pm 0.9$ with an 
equivalent Hydrogen column density of 
$0.1 \pm ^{2.5}_{0.1} \times 10^{21}$~cm$^{-2}$ or if the column density is 
fixed to be $3\times 10^{21}$ cm$^{-2}$, the interstellar value in that 
direction predicted from the HI map by Dickey \& Lockman (1990), the photon 
index is $2.6 \pm 0.7 $. The bremsstralung and MEKAL 
models are poorly constrained with kT $>$ 0.7 keV and $>$ 1.7 keV 
respectively, even when the absorption is fixed at the interstellar value. 
Fitting together the MOS and PN spectra for this region give similar and 
consistent spectral fits. The summed MOS spectrum from the larger region
gives a power law photon index of $1.5 \pm 0.4$ and an absorption of 
$0.15 \pm _{0.11} ^{0.15} \times 10^{21}$ cm$^{-2}$ with an unabsorbed 
0.5--2 keV flux of $6 \times 10^{-14}$ \ergcms. If the absorption is 
fixed  at the interstellar value gives a photon index of $1.8 \pm 0.3$. 
The best fit bremsstralung model for the larger region gives a kT $>$ 4 keV. 
In Figure 4 the sum of the MOS spectra for the larger region is shown with 
the best fitting power law model for the inner region. There is a suggestion 
that the spectrum of the larger region has a high energy excess, perhaps 
from contaminating unresolved sources.

\subsection{Bright Serendipitous Sources}

There are many faint sources in the field, that previously would have been 
hard to detect when the bright X-ray binary was active, because of 
scattering in the telescopes. 
Three of these field sources stand out in the iron band image and 
these are indicated with circles in Figure 2. To investigate these 
sources we extracted 
PN spectra (because this detector has a harder response).
The three spectra are shown in Figure 5. They are quite different, with 
source 1 (R.A.(J2000)=$ 17^{h} 59^{m} 00^{s}.8$ 
Dec(J2000)=$-33^{o} 45'' 48'.5$) displaying a prominent iron 
K line, source 2 (R.A.(J2000)= $17^{h} 57^{m} 58^{s}.7$  
Dec(J2000)=$-33^{o} 46' 26''.5$ ) a steeper featureless 
spectrum and source 3 (R.A.(J2000)= $17^{h} 59^{m} 21{^s}.6$ 
Dec(J2000)= $-33^{o} 53' 14''.3$) showing strong 
absorption.  We summarize the spectral fits to a power law model in 
Table 2. 

Three sources were previously detected from ROSAT observations, when 
4U1755-33 was active and they are marked in Fig 2 with squares. 
These three ROSAT sources were the brightest sources after  4U1755-33 reported in the 
WGACAT ROSAT source catalog for this field (1WGA J1758.6-3341,  
1WGA J1758.8-3341 and 1WGA J1759.3-3350).  
There were about eight other fainter detections reported in WGACAT that do 
not appear in the XMM-Newton image -- we ascribed these to false detections 
generated by the scattered light from  
4U1755-33.

Two of the three new "iron band" sources in the XMM-Newton image 
are absorbed and would not have been detected by ROSAT which had a soft $<$ 2 keV spectrum. 
Source 2 should have been detected by ROSAT and examining the ROSAT PSPC 
field there is a faint source at this position with a 0.1--2 keV band 
count rate of $\sim$ 0.005 ct/s, consistent with that predicted. 

The {\it ASCA}  source counts (e.g. Kushino \etal\  2002) predict in the 2--10 keV 
band 10 sources per square degree $> 10^{-13}$ \ergcms. 
The XMM-Newton field of view covers $\sim$ 0.25 square arc min, 
so 2--3 extragalactic sources above this flux is consistent the observation 
of 3 sources. While perhaps one of these sources might be galactic, 
it seems likely that at least two and probably all three are background AGN.

\section{A Fossil Jet from 4U1755-33}

The XMM-Newton image of the region surrounding 4U1755-33  reveals a 7 arc 
minute long jet-like structure, centered on the position of the LMXB. This is very 
suggestive that the jet was ejected from 4U1755-33 when it was X-ray active. 
4U1755-33 is located at l=357$^o$ and b=-4.8$^o$, i.e. well off the galactic plane. 
The X-ray dipping seen from this source indicate it is observed close to the orbital 
plane and so the jets would be expected 
to be symmetric, as observed. The jet runs at an angle of $\sim 70^o$ to the 
galactic plane, so it is unlikely to be associated with the galactic ridge. 

Another possibility is that the emission is a scattering halo from  dust 
intervening between the source. To produce such a narrow 
ridge would require a rather unusual geometry for the intervening material.  
The spectrum of a scattering halo would have a spectral 
index steeper by unity than that of the source (Xu, McCray \& Kelley 1986). 
The spectrum of 4U1755-33 when it was active was very steep, with an 
equivalent photon index of order 4 (White {\it et al} 1984) -- the opposite 
of that observed in the extended emission. 

X-ray emission from jets has been observed by the  {\it Chandra} X-ray observatory 
from many black hole systems, both galactic and extra-galactic.
The spectra of these jets are typically power laws with a photon 
index ranging from $\sim$ 1.5 -- 2.2 (e.g. Kaaret \etal\ 2002; 
Siemiginowska et al 2002; Wilson and Yang 2002), suggesting that 
non-thermal processes dominate. However, the jet from SS433 shows line emission 
indicating a hot gas with a temperature of $10^7$ to $10^8$ K. This should adiabatically  
cool along the jet (e.g. Brinkmann {\it et al} 1991), which is not seen 
from the 4U1755-33 jet, although recent {\it Chandra} observations demonstrate 
reenergization of the SS433 jet maybe possible (Migliari, Fender \& Mendez 2002).

The quality of the 4U1755-33  jet spectra is not high enough to distinguish thermal 
or non-thermal models. When fit with a power law the 4U1755-33 jet has a photon index 
of 1.5 to 1.9, consistent with that measured from other galactic and extra-galactic jets.
 The knots in the 4U1755-33 jet are also very reminiscent of  similar features 
seen in other X-ray jets (e.g. Wilson \& Yang 2002). 
If 4U1755-33 is at a distance of 4 kpc, then a 3.5 arc min long jet corresponds to 4 pc. 
If the jet has a velocity of close to the speed of light, then it would take $\sim 13$ yr 
to expand to this length -- consistent with the $>$ 25 yr active phase of 4U1755-33. 
If the jet is no longer being fed since the source has been quiescent, then we might 
expect to see a hole developing centered on 4U1755-33 and  indeed there is 
some evidence  for  such a feature (Figure 1). 

\section{Discussion \& Conclusions}

The XMM-Newton observation reported here did not detect any quiescent emission 
from 4U1755-33. The upper limit to the source luminosity depends on the assumed 
spectrum and distance. If the source is at the lower end of the assumed ranges 
(a distance of 4 kpc and temperature of 2 keV) then the upper limit of 
$\sim 10^{31}$ \ergs\ is comparable to other black hole transients in quiescence 
with similar orbital periods (Lasota \& Hameury 1998; Narayan, Garcia \& 
McClintock 2002), which would strengthen the black hole candidacy of 4U1755-33.
 If the distance is as far as 9 kpc and the underlying spectrum $\sim$10 keV, 
then the upper limit to the luminosity of $\sim 10^{32}$ \ergs\ starts to 
overlap the area where both neutron stars and black holes accreting at low 
accretion rates are found (see Figure 4 in Lasota \&  Hameury 1998). 

The transition of 4U1755-33 into a quiescent state has allowed the surrounding
 region to be studied in detail.
A surprising result from this observation has been  
the detection of a faint extended X-ray jet that appears to have been ejected 
from 4U1755-33 . Jets are a common feature of black hole 
candidates in our galaxy and in AGN, and this result provides more circumstantial 
evidence that 4U1755-33 contains a black hole. 
The several parsec extent of the jet is consistent with a source that has been active 
for more than 25 yrs. Sensitive radio observations should be made to search for a radio counterpart to the X-ray jet.
Future X-ray observations will confirm the jet nature of this feature by observing 
the expansion of the knots and the central hole. This will provide a rare opportunity to test jet propagation models.

\section{Acknowledgements}

We acknowledge the XMM-Newton science survey center staff Mike Watson and 
Wolfgang Pietsch for their insight and suggestions regarding these data.

\clearpage

\section{References}

%\subsection*{References}
\begin{description}

\renewcommand{\itemsep}{0pt}

\item Brinkmann W., Kawai, N., Matsuoka, M. and
 Fink, H. H. 1991, A\&A, 241, 112.

\item Dickey, J. M. \& Lockman F. J., 1990, Ann. Rev. Ast. Astr., 28, 215.

\item Garcia, M.R., McClintock, J.E., Narayan, R., Callanan, P. and 
Murray, S.M. 2000, ApJ, 553, L47.

\item Geherls, N.,1986, ApJ, 303, 336.
  
\item Giacconi, R., Murray, S., Gursky, H., Kellogg, E., Schreier, E., Matilsky, T., Koch, D. and Tananbaum, H. 1974, ApJS,  27,  37.

\item Hameury, J.-M., Barret, D., Lasota, J.-P., McClintock, J.E., 
Menou, K., Motch, C., Olive, J.-F., and Webb, N., 2002, A\&A, in press.

\item Jansen, F., Lumb, D., Altieri, B., Clavel, J.,
 Ehle, M., Erd, C., Gabriel, C., Guainazzi, M.,
 Gondoin, P., Much, R., {\it et al} 2001, A\&A, 365, L1.

\item Kaaret P., Corbel, S., Tomsick J.A., Fender, R., Miller, J.M., 
Orosz, J.A.,  Tzioumis, A.K. and  Wijnands, R.  2002, astro-ph/0210401.

\item Kushino, A., Ishisaki, Y., Morita, U., Yamasaki, Y., Ishida, M., Ohashi, T., and Ueda, Y. 2002, PASJ, 54, 327

\item Lasota, J.-P. and Hameury, J.-M, 1998, proceeding of 
``Accretion Processes in Astrophysical Systems: Some Like it Hot!'', 
College Park, MD, October 1997, eds S. S. Holt and T. R. Kallman, 
AIP Conference Proceedings 431., 351.

\item Mason, K.O., Parmar, A.N. \& White, N.E. 1985, MNRAS, 216, 1033.

\item Menou, K., Esin, A.A., Narayan, R., Garcia, M.R., Lasota, J.-P., and McClintock, J.E. 1999, \apj 520 276

\item Migliari, S., Fender, R., and Mendez, M. 2002, Science 297, 1673.

\item Narayan, R., McClintock, J.E., \& Yi, I., 1996, ApJ, 457, 821.

\item Narayan, R., Garcia, M.R., \& McClintock, J.E. 2002. Proc. IX 
Marcel Grossman Meeting, Eds V. Gurzadyan, R. Jantzen and R. Ruffini, 
Singapore: World Scientific Press, astro-ph/0107387.

\item Pan, H.C., Skinner, G. K., Sunyaev, R. A. and
 Borozdin, K. N.  1995, MNRAS, 274, 15P.

\item Roberts, M. S. E., Michelson P. F., Cominsky, L. R., Marshall F. E. , 
Corbet R. H. D., and Smith E. A. 1996, IAUC 6302.

\item Seon, K., Min, K., Yoshida, K., Makino, F., van der Klis, M., 
van Paradijs, J., and Lewin, W. 1995, ApJ, 454, 463.

\item Siemiginowska A., Bechtold, J., Aldcroft, T. L., Elvis, M.,
 Harris, D. E. and Dobrzycki, A. 2002 570, 543-556

\item Wilson A. S.,  and Yang Y. 2002, ApJ, 568, 133.

\item Wachter, S. and Smale, A.P., 1998, ApJ, 496, 21.

\item White, N.E. and  Marshall, F.E., 1984, ApJ, 281, 354.

\item White, N.E.,  Parmar, A. N., Sztajno, M., Zimmermann, H. U.,
 Mason, K. O. and Kahn, S. M. 1984, ApJ, 283, L9.

\item Xu, Y., McCray, R., and Kelley, R. 1986, Nature 319, 652.

\end{description}

\clearpage

\begin{table}

\begin{center}

\begin{tabular}{lccccc}  

\multicolumn{6}{c}{Table 1:  Flux \& Luminosity upper limits}\\  
\tableline \tableline\\
    & &   \multicolumn{2}{c}{Flux}   &  
\multicolumn{2}{c}{Luminosity} \\ 
\cline{3-4} \cline{5-6} \\ 
Instrument & Temperature & 0.5-2 &        2-10  & 0.5-2 &        2-10\\
 & keV & \ergcms &        \ergcms  & \ergs &        \ergs\\
\tableline\\

MOS   & 2&   $9.0 \times 10^{-15}$ & $1.2\times 10^{-14}$ & $1.7\times 10^{31} d_{4kpc}^2$ & $2.3 \times 10^{31}d_{4kpc}^2$   \\
PN    & 2 &    $1.2\times 10^{-14}$ & $1.8 \times 10^{-14}$ & $2.3 \times10^{31}d_{4kpc}^2$ & $3.5 \times 10^{31}d_{4kpc}^2$   \\

MOS   & 10 &  $9.0  \times 10^{-15}$ & $1.9 \times 10^{-14}$ &  $1.7 \times 10^{31}d_{4kpc}^2$  &       $3.6 \times 10^{31}d_{4kpc}^2$   \\
PN    & 10  &  $1.2 \times 10^{-14}$ & $2.6 \times 10^{-14}$ &  $2.2 \times 10^{31}d_{4kpc}^2$   &       $5.0 \times 10^{31}d_{4kpc}^2$ \\

\tableline
\end{tabular}
\end{center}
\end{table}

\clearpage

\begin{table}

\begin{center}

\begin{tabular}{lccccccc}  

\multicolumn{8}{c}{Table 2: Spectral Parameters of the Serendipitous XMM-Newton Sources}\\

\tableline \tableline\\
   & & &   \multicolumn{3}{c}{Absorbed Flux} & \multicolumn{2}{c}{Line}  \\
\cline{4-6} \cline{7-8} \\ 

Source & Index & N$_{H}$ & 0.5-2 & 2-10 & 5-10 & Energy &  EQW \\
 & & $ 10^{22}$~cm$^{-2}$ & \ergcms & \ergcms & \ergcms & keV & eV \\
\tableline\\

1 & $1.71 \pm ^{0.27} _{0.23}$ & $0.96 \pm ^{0.37} _{0.25}$ & 
$4 \times 10^{-14}$ & $3.17  \times 10^{-13}$ & $ 1.8 \times 10^{-13}$ & 
$6.20 \pm ^{0.2} _{0.16}$ & 876 \\ 

2 & $2.08 \pm 0.27 $ & $0.32 \pm ^{0.13} _{0.07}$ & 
$5.6 \times 10^{-14}$ &$ 1.26  \times 10^{-13}$ & $ 5.4 \times 10^{-13}$ & & \\

3 & 1.7 $^a$ & $13.7 \pm ^{5.8} _{4.11}$ & 
$4.8 \times 10^{-17}$ & $2.8  \times 10^{-13}$ & $ 2.2 \times 10^{-13}$ &  &  \\ 
\tableline
\end{tabular}
\end{center}
$^a$ Power law index keept fixed
\end{table}

\clearpage
\small
\begin{figure}
\figurenum{1}                            
\centerline{\includegraphics[scale=0.5]{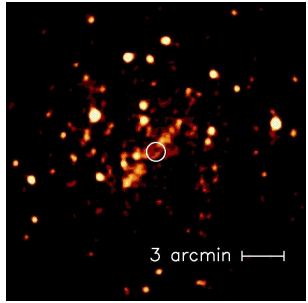}}
\figcaption{XMM-Newton 21 x 21 arc min image  of the 4U1755-338 region using the MOS-1, MOS-2 and PN summed together in the energy band 0.5--4.5 keV. 
The location of 4U1755-338 is at the center of the image and marked 
with a circle. The image has been 
smoothed using a gaussian with an 8 arcsec $\sigma$.}
\end{figure}

\begin{figure}
\figurenum{2}                            
\centerline{\includegraphics[scale=0.5]{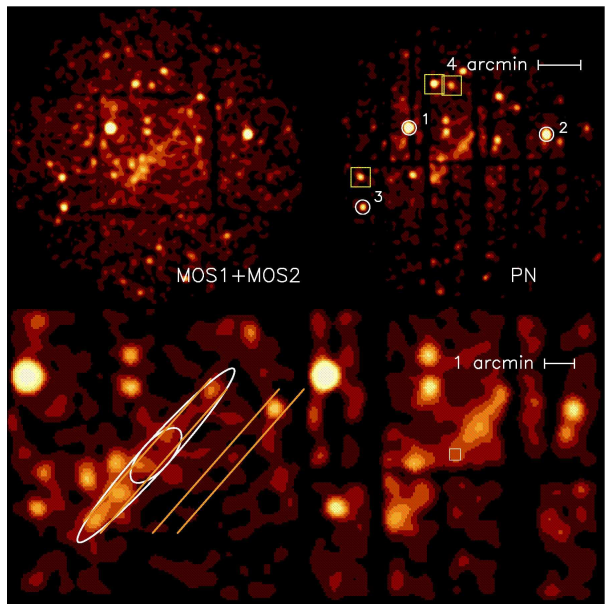}}
\figcaption{XMM-Newton images of the 4U1755-338 region in the 0.5--10 keV 
energy band. The upper panel shows the entire field of view of the MOS (left)
and the PN (right). The jet is evident in the both instruments. The sources marked in the PN image
with circles are serendipitous sources that stand out in the 6.2--6.8 keV energy band. Those marked with squares are the ROSAT WGA sources detected by XMM-Newton. The lower panel shows the central 10 arc min of the FOV of the MOS (left)
and PN (right). In the MOS image are drawn the regions (ellipses) from
where spectra were extracted and the regions (parallel lines) from the 
intensity profiles, shown in Fig 3, were extracted. The square in the
lower right image at the image center is drawn at the position of 4U1755-33.}
\end{figure}

\begin{figure}
\figurenum{3}                            
\centerline{\includegraphics[scale=0.5,angle=-90]{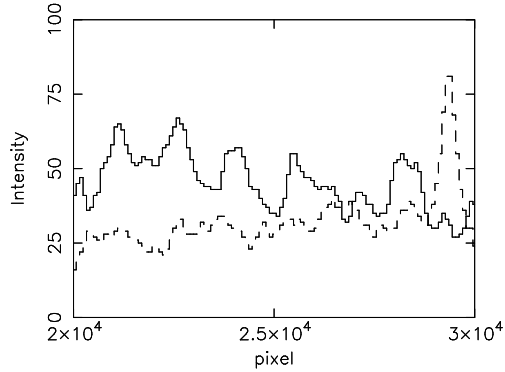}}
\figcaption{The intensity profile obtained from the region marked with parallel
lines in the lower left of the Figure 2 in the 0.5--10 keV energy range.
The intensity was obtaned from a image smoothed with a $\sigma$ of 6 arcsec.
The dash line represents the region that includes a point source.}
\end{figure}

\begin{figure}
\figurenum{4}                            
\centerline{\includegraphics[scale=0.5,angle=-90]{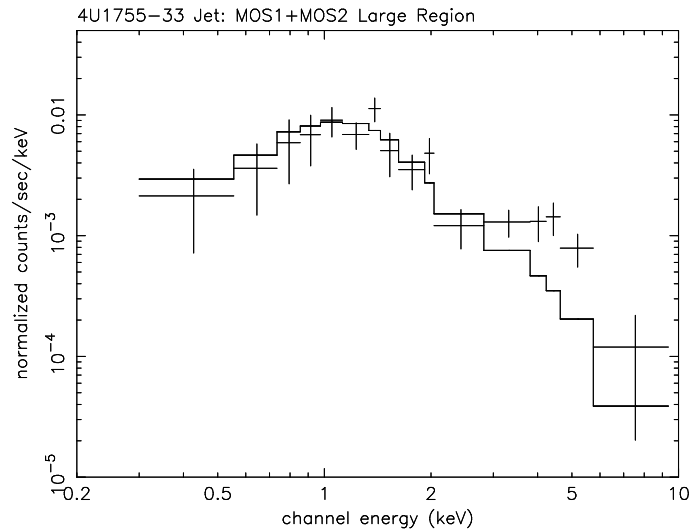}}
\figcaption{The sum of the two MOS spectra. The spectra were extracted from 
the ellipses that includes the entire jet and are overlayed
with the best model spectrum obtained from the central region.}
\end{figure}

\begin{figure}
\figurenum{5}                            
\centerline{\includegraphics[scale=0.5,angle=-90]{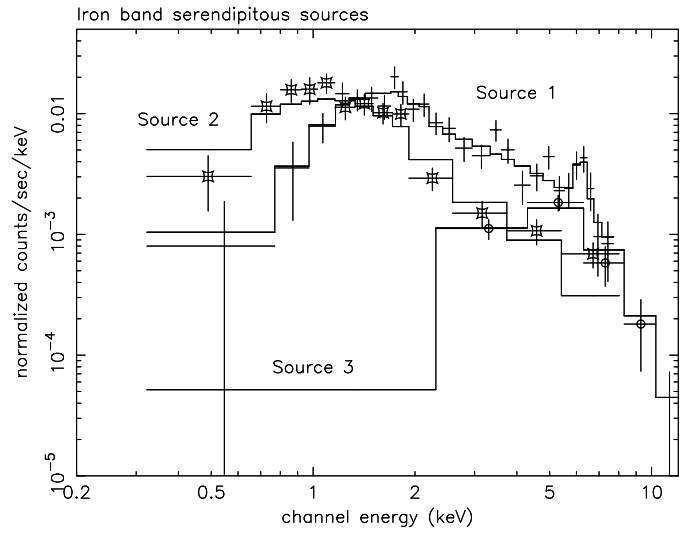}}
\caption{PN spectra extracted from the three serendipitous sources 
detected in the 6.2--6.8 keV region. The spectra are quite different 
with Source 3 very absorbed and Source 1 showing an iron line.}
\end{figure}

\end{document}